\documentclass[sigconf]{acmart} %,

\pdfoutput=1

\usepackage[export]{adjustbox}
\usepackage{enumitem}
\usepackage{float}
\usepackage{dblfloatfix}
\RequirePackage{amsfonts,%
    algorithmic,%
    textcomp,%
    xcolor,%
    framed,%
    fontawesome5%
}
\AtBeginDocument{%
  \providecommand\BibTeX{{%
    \normalfont B\kern-0.5em{\scshape i\kern-0.25em b}\kern-0.8em\TeX}}}

\acmConference[ASE]{Automated Software Engineering}{October 27 -- November 01,
  2024}{Sacramento, CA, USA}

\begin{document}

\title{Towards Extracting Ethical Concerns-related Software Requirements from App Reviews
}

\author{Aakash Sorathiya}
\affiliation{%
  \institution{University of Calgary}
  \city{Calgary}
  \country{Canada}}
\email{aakash.sorathiya@ucalgary.ca}

\author{Gouri Ginde}
\affiliation{%
  \institution{University of Calgary}
  \city{Calgary}
  \country{Canada}}
\email{gouri.deshpande@ucalgary.ca}

\begin{abstract}
As mobile applications become increasingly integral to our daily lives, concerns about ethics have grown drastically. Users share their experiences, report bugs, and request new features in application reviews, often highlighting safety, privacy, and accountability concerns. Approaches using machine learning techniques have been used in the past to identify these ethical concerns. However, understanding the underlying reasons behind them and extracting requirements that could address these concerns is crucial for safer software solution development. Thus, we propose a novel approach that leverages a knowledge graph (KG) model to extract software requirements from app reviews, capturing contextual data related to ethical concerns. Our framework consists of three main components: developing an ontology with relevant entities and relations, extracting key entities from app reviews, and creating connections between them. This study analyzes app reviews of the Uber mobile application (a popular taxi/ride app) and presents the preliminary results from the proposed solution. Initial results show that KG can effectively capture contextual data related to software ethical concerns, the underlying reasons behind these concerns, and the corresponding potential requirements.
\end{abstract}


\maketitle

\section{Introduction} \label{introduction}
The development of trustworthy and ethical software requires the prioritization of user experience. User reviews on app stores and online platforms offer a rich tapestry of real-world experiences. However, extracting and analyzing the ethical concerns embedded within them remains a significant challenge.

Current research on user feedback in software engineering focuses on user sentiment and feature requests \cite{genc2017systematic} \cite{panichella2015can} \cite{tavakoli2018extracting} \cite{guzman2014users}. While this research provides valuable insights into user preferences and areas for improvement, it does not capture the nuanced ethical dimensions that trouble users. Understanding these concerns is crucial for building software that aligns with user values, fosters trust, and avoids unintended biases. 

\textbf{Motivating example:} Several studies have begun to explore user perspectives on ethical issues within the software. Research by Besmer et al. \cite{besmer2020investigating} and Nema et al. \cite{nema2022analyzing} highlights users’ concerns regarding privacy violations and data security practices within mobile apps. Discriminatory algorithms and the potential for bias within software functionalities are also emerging areas of concern, as evidenced by the work of Tushev et al. \cite{tushev2020digital} and Olson et al. \cite{olson2023along}. Manipulative design patterns that pressure users or exploit psychological vulnerabilities are also a growing concern, as identified by Olson et al. \cite{olson2024best}. However, these approaches often struggle to comprehensively capture and categorize the diverse ethical concerns expressed within app reviews. Research \cite{olson2023along} and \cite{olson2024best} have used classification techniques to filter and categorize user feedback by their predominant ethical concerns, but these techniques lack contextual information.

For instance, consider the case of a ride-sharing application where User A books a ride on Uber, and it is accepted by Driver A, who is registered with Uber. However, instead of Driver A, the request is serviced by Driver B, a proxy driver for Driver A. This practice creates a serious safety issue for app users, which is highlighted by one of the users in the application reviews, as shown below. However, previous studies classify this review as a "safety" ethical concern, which is insufficient to understand its underlying reason and the requirement to address it.
\\
\hfill\\
\begin{tabular}{||p{8cm}}
\faIcon[regular]{thumbs-down} \textbf{App Review (Uber):} Uber app must have a provision to load the driver's pic before starting of trip so that Uber runs a face recognition match before start of trip. This is to ensure that proxy drivers don't drive cars making it unsafe for passengers. At least this feature can be enabled for late night rides enhancing security of female passengers.
\end{tabular}
\\

\par \textbf{To address this research gap}, we propose an approach that leverages a KG and seeks to answer the research question:\textbf{ \emph{How effective is knowledge graph to visualize and capture the underlying reasons and requirements related to the ethical concerns expressed by users in the application reviews?}} 

\noindent \textbf{Rationale:} KGs are powerful tools for capturing and representing relationships between entities and concepts \cite{googlekg}, widely used in various domains such as healthcare, e-commerce, and finance \cite{gu2017visualizing}, \cite{abu2021domain}. In software engineering, KG has been used for software vulnerability mining, security testing, and API recommendation \cite{wang2023application}. These applications of KG highlight its capabilities in knowledge reasoning, information retrieval, information mining, and their related dependencies.

Thus, taking inspiration from various examples, this research explores the utility of KG in capturing ethical concerns and their underlying reasons, thereby extracting the requirements from app reviews.
\\
\noindent
\textbf{ Our contributions in this study are twofold:} We propose (1) An approach to comprehensively understand the reasons behind users' ethical concerns and extract corresponding software requirements, and (2) Visualize and understand the interconnection between various ethical concerns.

\begin{figure}
    \centerline{\includegraphics[scale=0.18]{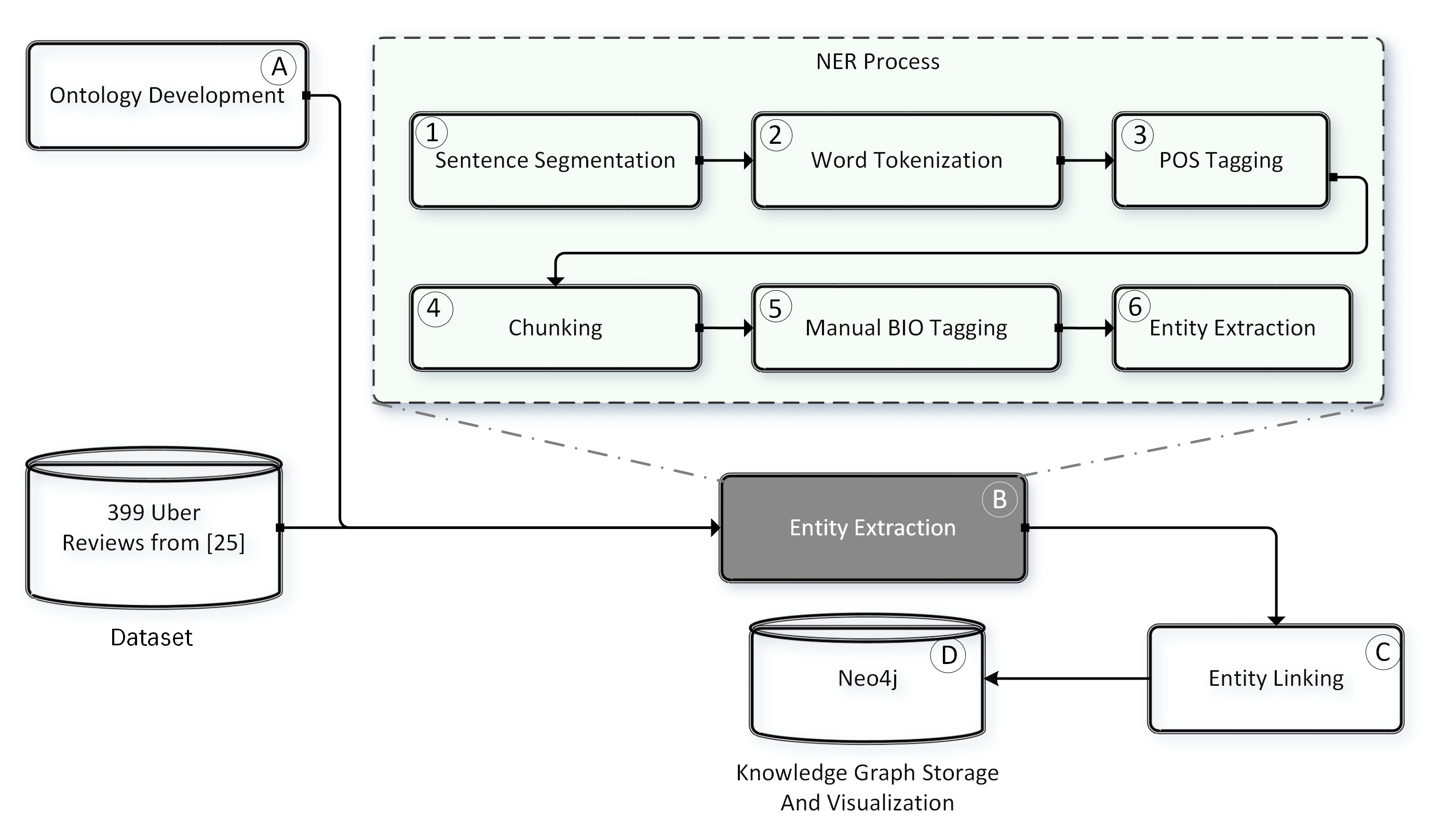}}
    \caption{Our study design. Entity extraction is performed based on the ontology, followed by entity linking for KG construction. KG is stored in the graph database for incremental expansion of the dataset}
    \label{fig:design}
\end{figure}

\section{Our approach} \label{methodology}
 We propose a novel framework to capture the ethical concerns expressed by users in app reviews. Figure~\ref{fig:design} shows an overview of our study design which consists of three main modules: A) collection of app reviews B) construction of the KG, and C) storage and visualization of the KG.
\begin{enumerate}[leftmargin=*, label=(\Alph*)]
\item \textbf{App reviews:}
In our study, we use the dataset and ethical concerns identified in the previous work by Olson et al.\cite{olson2024best}. This dataset consists of 3,101 manually labeled Google Play Store reviews from ten popular apps: TikTok, Uber, Facebook, YouTube, Instagram, Linkedin, Vinted, Zoom, Alexa, and Google Home. The manually annotated labels indicate if a review contains an ethical concern and its type. Our study uses the Uber app's reviews, which account for 399 reviews. Table~\ref{tab:conerns} shows the types of ethical concerns identified for the Uber app and their percentage distribution.

\begin{table}[!htbp]
\centering
    \caption{Percentage-wise distribution of ethical concern types identified from reviews for Uber application by Olson et al. \cite{olson2024best}.}
    \begin{center}
        \begin{tabular}{ll|ll}
       
            Safety& 27.1\% & Accessibility & 2.8\% \\
            \hline
            Accountability& 17.3\%& Sustainability& 1.5\% \\
            \hline
            Scam& 15.8\%& Identity Theft& 1.5\% \\
            \hline
            Discrimination& 7\%& Cyberbullying/Toxicity& 0.7\% \\
            \hline
            Transparency& 4.5\%& Spreading False Information& 0.5\% \\
            \hline
            Privacy& 3.3\%& Inappropriate Content& 0.2\% \\
         
        \end{tabular}
        \label{tab:conerns}
    \end{center}
\end{table}
\begin{figure*}[!htpb]
\begin{minipage}{.47\textwidth}
    \centering
   \includegraphics[valign=t, scale=0.17]{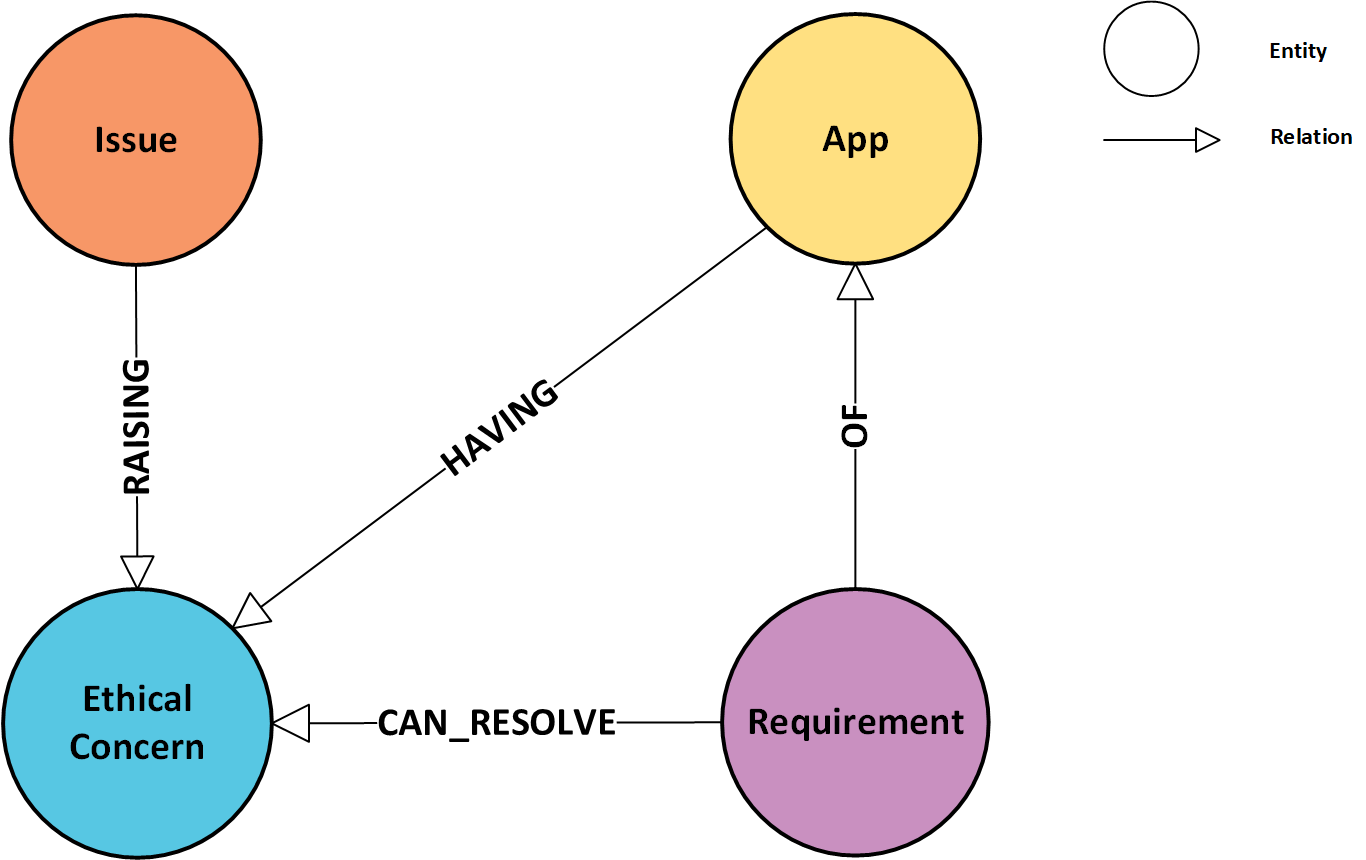}
    \caption{This Ontology graph shows the entities and relations identified during ontology development. Entities (Table \ref{tab:ed}) are represented by circles and relations by arrows.}
    \label{fig:ont}
    \end{minipage}
   \hfill
    \begin{minipage}{.47\textwidth}
    \centering
    \includegraphics[valign=t, scale=0.6]{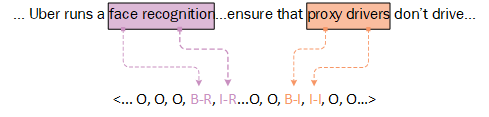}
 
    \caption{An example of BIO-tagged review. \textit{O}: non-entities, \textit{B-R}: first word of the Requirement entity, and \textit{I-R}: All subsequent words. \textit{EC}: EthicalConcern entity \& \textit{I}: Issue entity.}
    \label{fig:bio}
    \end{minipage}
\end{figure*}

\item \textbf{Knowledge graph construction:}
We follow the development process by Tamašauskaitė et al. \cite{tamavsauskaite2023defining} to construct a KG. We develop the domain ontology to extract the entities from app reviews and create relationships based on the ontology.

\textbf{Ontology Development:}
Before building a KG, we need to define an ontology to store the knowledge required. The process outlined by Bravo et al. \cite{bravo2019methodology} is used for ontology development, and below are the steps followed for this development.

\begin{itemize}[leftmargin=*]
    \item \textit{Ontology Requirements Specification}:
This ontology aims to support software developers and researchers in discovering the contextual information of ethical concerns from app reviews and help enhance software design by considering the requirements to address these concerns effectively. Thus, we designed the following competency questions based on our use case.
\begin{itemize}
    \item What are the ethical concerns raised by users?
    \item What are the underlying reasons behind these concerns?
    \item What requirements can be identified from app reviews to address these concerns?
    \item Are there any common patterns between various ethical concerns across applications of the same genre?
\end{itemize}

\item \emph{Ontology Design: } In this step, we identify the set of terms or concepts (listed in Table~\ref{tab:ed}) based on the competency questions above.
\begin{table}[b]
    \centering
    \caption{Definition of entities identified during ontology development.}
    \begin{tabular}{p{1.5cm}  p{6.3cm}}
         \textbf{Entity} & \textbf{Definition} \\
         \hline
         App & Name of the application (Ex: Uber) \\
         \hline
         Issue & An issue faced by users which is the underlying reason for the rise of ethical concern. \\
         \hline
         Ethical Concern & An ethical concern reported in app reviews. \\
         \hline
         Requirement & A new feature/functionality suggested by users that might address the ethical concerns. \\
        
    \end{tabular}
    \label{tab:ed}
\end{table}
    
\item\emph{Ontology Construction}: In this step, the relations between the concepts introduced in the previous step are defined, and the graph is then constructed using the Protégé editor \cite{DBLP:journals/aimatters/Musen15}, as shown in Figure~\ref{fig:ont}.
\end{itemize}

\noindent
\textbf{Entity Extraction:}
Entity extraction aims to discover and detect entities in a wide range of data. One of the most frequently applied methods is named entity recognition (NER), which focuses on discovering and classifying entities to the predefined categories or types \cite{tamavsauskaite2023defining}. 

The NER process in our study design involves the manual labeling of input data. First, we segment the raw feedback into sentences and divide it into words and phrases. Part-of-speech tagging is then used to assign POS tags to each word, followed by a chunking operation to group these tokens into chunks. After this, the entities are manually labeled in two steps with the BIO tags scheme mentioned by \cite{ratinov2009design}. The BIO tags scheme identifies the \textbf{B}eginning, the \textbf{I}nside, and the \textbf{O}utside of the text segments. In the first step, the first author labeled the entities with BIO tags in an iterative process. After each iteration, the second author cross-checked the labeled entities, and any inconsistent tagged data was further relabeled.

Finally, the labeled entities are extracted from the processed data. Steps 1 to 4 were carried out using the NLTK library \cite{loper2002nltk}. The "app" entity is directly extracted from the data, as the study focuses only on the "Uber" application. Figure~\ref{fig:bio} shows an example of the BIO-tagged review. \textbf{R} is the abbreviation for Requirement entity, \textbf{I} is Issue entity, and \textbf{EC} means EthicalConcern entity. \\
\noindent
\textbf{Entity Linking:}
Extracted entities must be linked using the defined relations. For structured data and defined ontology, relations are explicit and easily identifiable \cite{tamavsauskaite2023defining}. Hence, as shown in Figure~\ref{fig:ont}, we define four types of relations and create links between the extracted entities based on the ontology. An example of such a link is: \emph{Uber (App) – HAVING – Safety (EthicalConcern)}.

\item \textbf{Knowledge graph storage and visualization:}
We utilize the graph database Neo4j to store and manage the extracted data due to its efficient handling of interconnected relationships \cite{fernandes2018graph}, which can provide deeper insights through node-link patterns. 
% Figure~\ref{fig:graph} shows a small part of this network visualization of app reviews.
\end{enumerate} 
\begin{figure*}[!htpb]
    \centering\centerline{\includegraphics[scale=0.23]{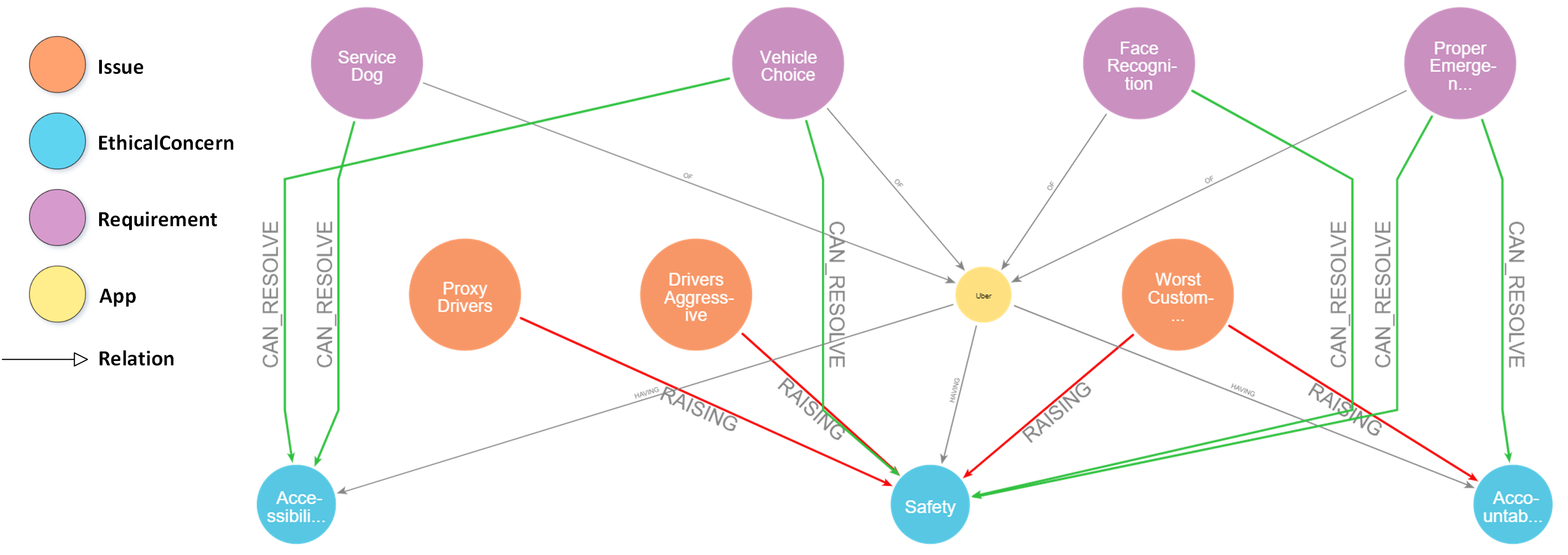}}
    \caption{An example of a KG generated in Neo4j from the extracted data. The graph shows the visualization of the scenarios as described in Section~\ref{results}. \textcolor{red}{Red} arrows represent the underlying reasons for the ethical concerns, and \textcolor{green}{green} arrows show the potential requirements suggested by users that can address the ethical concerns.}
    \label{fig:graph}
    % \vspace{-5mm}
\end{figure*}

\section{Preliminary findings} \label{results}
The generated KG had 467 nodes and 544 relations from which we could extract 14 ethical concerns-related requirements. 
Figure \ref{fig:graph} shows a small subset of KG from Neo4j  from our proposed approach. In this example, we have identified three EthicalConcern entities: \textit{Accessibility}, \textit{Safety}, and \textit{Accountability}. To address these concerns, our approach has extracted four specific Requirement entities: \textit{Service Dog}, \textit{Vehicale Choice}, \textit{Face Recofnition}, and \textit{Proper Emergency Number}. Our approach also identified three Issue entities that are the underlying reasons why these ethical concerns arise: \textit{Proxy Drivers}, \textit{Drivers Aggressive}, and \textit{Worst Customer Support}. To further illustrate our approach, in \textbf{a)}, we explain this network using the two example app reviews, and in  \textbf{b)} we explain the interconnection of various ethical concerns.
\newline
\noindent \textbf{a) Reasons and Ethical Concerns-related Requirements:} \\ 
\textbf{App Review 1}: \emph{Uber app must have a provision to load the driver's pic before starting of trip so that Uber runs a face recognition match before start of trip. This is to ensure that proxy drivers don't drive cars making it unsafe for passengers. At least this feature can be enabled for late night rides enhancing security of female passengers.}
\\
Our KG analysis uncovers that this review highlights a "Safety" concern, specifically related to the "Proxy Drivers" issue. The reviewer expresses concern about the potential risks associated with proxy drivers operating on the Uber platform. Moreover, the review suggests including "Face Recognition" technology in the app to help verify the identity of drivers and ensure a safer experience for users.
\\
\noindent \textbf{App Review 2}: \emph{There is no customer support for uber. Safety number does not provide any support. Its v unsafe. Big company without basic safety precaution is unacceptable. Changing route they are charging such high charges(1.2x). Already fares are higher on top of it these surge is v bad. That too in corona situation there is no traffic at all. First thing there should be proper customer support, proper emergency number. Actual person should be there in call centre instead of machine to answer calls.}
\\
Our KG analysis reveals that this review is centered around “Safety” concerns and highlights the issue with Uber’s customer service.  Specifically, the reviewer expresses dissatisfaction with the company's support system and suggests it should have reliable emergency numbers in place to ensure a safer experience for users. \\ \\ 
\noindent \textbf{b) Interconnection of Ethical Concerns:}
Our KG enables us to visualize the intricate relationships between various ethical concerns,
allowing us to identify common reasons for the rise of ethical concerns and potential requirements to address them. As illustrated in Figure \ref{fig:graph},\textit{Worst Customer Support} issue is directly linked to both \textit{Safety} and \textit{Accountability} concerns, and requirement \textit{Proper Emergency Number} could help resolve these interconnected ethical concerns. This connection highlights the importance of providing accountable customer support during emergencies. By acknowledging these interdependencies, we can pinpoint areas that require improvement and extract requirements that could help mitigate ethical concerns.
 
\section{Related Work} \label{relatedwork}
This section summarizes relevant prior research on app reviews, their classification, sentiment analysis, implications, and KGs related to app reviews. 

\textbf{App Reviews}: App reviews are a valuable source of information for understanding user experiences. Research has leveraged these reviews to classify bug reports, feature requests, and user praise, prioritizing improvements and addressing user concerns \cite{maalej2015bug} \cite{panichella2015can}. Additionally, the research explores trends and implications within the mobile app review landscape, providing insights into user behavior and app store dynamics \cite{hoon2013analysis} \cite{martin2016survey}. App reviews can also be used to recommend improvements \cite{pagano2013user}, identify informative reviews for developers \cite{chen2014ar}, and guide release planning based on user sentiment \cite{villarroel2016release}. Fine-grained sentiment analysis of app reviews helps developers understand specific feature perceptions, informing decisions about future development \cite{guzman2014users} \cite{gu2015parts}. Moreover, app reviews can be used to understand user needs, identify desired functionalities \cite{iacob2013retrieving}, and inform software requirements engineering processes \cite{carreno2013analysis}. While prior research has extensively explored user feedback in software engineering, the ethical dimensions within app reviews remain relatively nascent, with some recent studies examining specific ethical concerns raised by users \cite{olson2023along} \cite{olson2024best}.

\textbf{Knowledge Graph}: Large-scale, high-quality, general-purpose knowledge bases (KBs) like Freebase \cite{bollacker2008freebase}, YAGO \cite{suchanek2008yago}, and DBpedia \cite{bizer2009dbpedia} have evolved since the concept of a KG was introduced by Google \cite{googlekg}. Industry-specific KGs are being researched to serve professional domains better. In the domain of app reviews, various ontologies, and knowledge bases have been developed to allow knowledge management and knowledge sharing. Morales-Ramirez et al. \cite{morales2015ontology} proposes a novel user feedback ontology founded on a Unified Foundational Ontology (UFO) that supports the description of analysis processes of user feedback in software engineering. Kifetew et al. \cite{channa2019requirements} examine mobile app ratings to extract requirements and assess app quality using interleaved and independent rating schemas, proposing an ontological strategy for accurate analysis and rigorous requirement extraction to satisfy users better. These review KGs are mainly used for decision support in the software evolution process and lack the consideration of ethical concerns in app reviews.
\vspace{-3mm}
\section{Conclusion and Future Plan} \label{future}
In this study, we presented a novel approach by utilizing NER and KG to perform contextual analysis of app reviews through ethical lenses. We used ride-sharing app reviews as an example to demonstrate our methodology’s application for ethics-related requirements extraction. 
Our preliminary results, through contextual analysis of app reviews in the ethical landscape, indicate some promising direction for extracting ethical concerns-related requirements. However, additional steps are needed to improve our approach further, which are listed below.  
\begin{itemize}[leftmargin=*]
    \item To capture more relevant entities and relations from the app reviews, we plan to refine the ontology using association rule mining \cite{volker2011statistical}.  We will also interview developers to understand and integrate their requirements into the ontology. This will help further understand the rise and mitigation of ethical concerns and their intricate relationships. 
    \item  We will utilize reviews of other similar applications from different app stores and social media platforms to enrich the knowledge base and gain insights into the interconnectedness of various ethical concerns across application genres for ethical consideration-related guidelines recommendations.
    \item We plan to build a fully automated framework to streamline the data collection from different sources, entity extraction, and entity linking. This will help to process the large corpus of app reviews dataset automatically. To automate the NER process, we plan to explore different machine learning models such as CRF \cite{lafferty2001conditional} and BI-LSTM-CRF \cite{huang2015bidirectional} 
    \item We will utilize the enriched KG to identify new ethical concerns that may not be explicitly mentioned or explored in the current literature. This idea is motivated by the concept of novel class discovery in the open-world machine learning \cite{zhu2024open}. 
    \item We envision designing a user-friendly tool to enable developers to visualize and derive the requirements that can help mitigate ethical concerns. This tool will also aid application users in understanding and comparing different applications from an ethical standpoint and deciding whether to use them.
\end{itemize}

\bibliographystyle{ACM-Reference-Format}
\bibliography{ref}

\end{document}